\documentclass[11pt]{article}

\usepackage{xcolor}
\usepackage{microtype}
\usepackage{url}            
\usepackage{amsfonts}       
\usepackage{nicefrac}       
\usepackage{mathtools, nccmath}
\usepackage{minitoc}
\usepackage{cancel}
\definecolor{darkgreen}{RGB}{0, 153, 0}

\usepackage{microtype}
\usepackage{graphicx}
\usepackage{booktabs} 

\usepackage{comment}
\usepackage{url}
\usepackage{pythonhighlight}

\usepackage{subfigure}
\usepackage{amsfonts}       
\usepackage{amsmath}
\usepackage{amsfonts}
\usepackage{comment}

\usepackage{amsthm}
\usepackage{amscd}
\usepackage{t1enc}
\usepackage{enumerate}
\usepackage{indentfirst}
\usepackage{listings}
\usepackage{graphicx}
\usepackage{graphics}
\usepackage{pict2e}
\usepackage{epic}
\usepackage{epstopdf} 
\usepackage{listings}
\usepackage{minitoc}

\newcommand{\bE}{\mathbb{E}}
\newcommand{\bP}{\mathbb{P}}
\newcommand{\bI}{\mathbb{I}}

\renewcommand{\nu}{\vartheta}

\usepackage{pgfplots} 

\usepackage{bm}
\usepackage{wrapfig}

\newtheorem*{theorem*}{Theorem}

\newtheorem{proposition}{Proposition}
\newtheorem{definition}{Definition}

\theoremstyle{definition}
\usepackage{algorithm}
\usepackage{algorithmic}

\usepackage{environ}
\usepackage{tikz}
\usepackage{float}

\newcounter{todos}

\newcommand{\E}{\mathbb{E}}

\usepackage{float}

\usepackage{xspace}
\newcommand{\coderec}{Copilot\xspace}

\usepackage{fontawesome}

\usepackage{paralist}

\usepackage{blindtext}

\usepackage[left=1in, right =1in, top=1in, bottom=0.7in]{geometry}

\pdfoutput=1
\usepackage{booktabs}
\usepackage{pgfplots}
\usepackage{graphicx}
\usepackage{tikz}
\usepackage{amsmath, amsthm}
\usepackage{amssymb}
\usepackage{graphicx}
\usepackage{setspace}
\usepackage{xfrac}
\usepackage{xstring}
\usepackage{xspace}
\usepackage{bm}
\usepackage{microtype}
\usepackage{graphicx}

\usepackage{booktabs} 

\usepackage{comment}
\pgfplotsset{compat=1.14}

\usepackage{gensymb}
\usepackage{amsfonts}       
\usepackage{nicefrac}       
\usepackage{microtype}      
\usepackage{amsmath}
\usepackage{amsfonts}
\usepackage{amssymb}
\usepackage{comment}
\usepackage[hidelinks]{hyperref}
\usepackage{amsthm}
\usepackage{amscd}
\usepackage{t1enc}
\usepackage{enumerate}
\usepackage{enumitem}
\usepackage[mathscr]{eucal}
\usepackage{indentfirst}
\usepackage{listings}
\usepackage{graphicx}
\usepackage{graphics}
\usepackage{pict2e}
\usepackage{epic}
\usepackage{epstopdf} 
\usepackage{wrapfig}
\usepackage{algorithm}
\usepackage{algorithmic}
\usepackage{amsthm}
\usepackage{tabularx}
\usepackage{float}
\usepackage{setspace}
\usepackage{fancyhdr}
\usepackage{paralist}
\usepackage{soul}
\usepackage{xcolor}

\definecolor{highlighter}{HTML}{fff100}
\sethlcolor{highlighter}
\usepackage[most]{tcolorbox}
\title{\textbf{When to Show a Suggestion? Integrating Human Feedback in AI-Assisted Programming}}
\allowdisplaybreaks
\usepackage{authblk}

\author[1]{Hussein Mozannar}
\author[2]{Gagan Bansal}
\author[2]{Adam Fourney}
\author[2]{Eric Horvitz}
\affil[1]{Massachusetts Institute of Technology, Cambridge, USA}
\affil[2]{Microsoft Research, Redmond, USA}
\date{}
\begin{document}
\newgeometry{left=1in, right =1in, top=0.6in, bottom=0.7in}

\begin{titlepage}
\maketitle
\vspace{-2cm}
\begin{abstract}
AI powered code-recommendation systems, such as Copilot and CodeWhisperer, provide code suggestions inside a programmer's environment (e.g., an IDE) with the aim of improving productivity. We pursue mechanisms for leveraging signals about programmers' acceptance and rejection of code suggestions to guide recommendations. We harness data drawn from interactions with GitHub Copilot, a system used by millions of programmers, to develop interventions that can save time for programmers. We introduce a utility-theoretic framework to drive decisions about suggestions to display versus withhold. The approach, conditional suggestion display from human feedback (CDHF), relies on a cascade of models that provide the likelihood that recommended code will be accepted. These likelihoods are used to selectively hide suggestions, reducing both latency and programmer verification time. Using data from 535 programmers, we perform a retrospective evaluation of CDHF and show that we can avoid displaying a significant fraction of suggestions that would have been rejected. We further demonstrate the importance of incorporating the programmer's latent unobserved state in decisions about when to display suggestions through an ablation study. Finally, we showcase how using suggestion acceptance as a reward signal for guiding the display of suggestions can lead to suggestions of reduced quality, indicating an unexpected pitfall.  
\end{abstract}

\begin{figure}[h]
	\centering
	\includegraphics[width=0.5\textwidth]{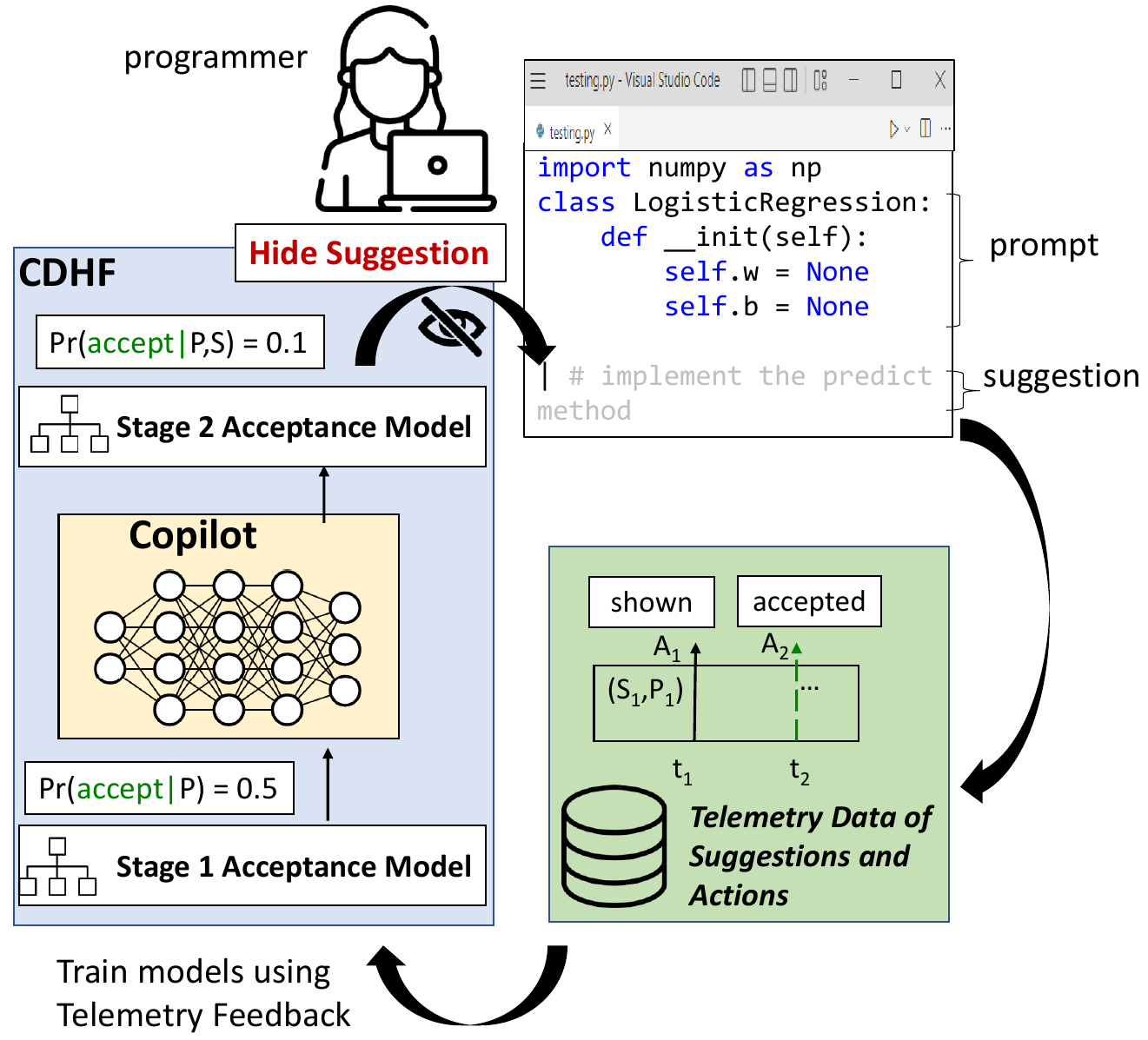}
     \vspace{-0.8em}

	\caption{ Operating mode of \coderec inside Visual Studio Code and how CDHF influences the interaction by selectively hiding certain suggestions. The data collected by the interaction is stored in telemetry and is used to train CDHF to create a feedback loop.}
	\label{fig:ide_coderec}
     \vspace{-1.5em}

\end{figure}

\end{titlepage}
\restoregeometry

\section{Introduction}

Code recommendation systems powered by large-scale neural language models, such as  Github Copilot \cite{copilot} and  Amazon CodeWhisperer \cite{codwhisperer}, are aimed at providing programmers with code suggestions to help improve their productivity. These systems usually operate by displaying the suggestion as {\em ghost text}---a grayed-out code suggestion inline inside the IDE. Programmers can accept the suggestion, browse through different suggestions, or reject the suggestion (see Figure~\ref{fig:ide_coderec}). The code suggestions appear either at the explicit invocation of the programmer or when the programmer pauses their cursor when writing code. GitHub reported a recent randomized study with 95 participants who wrote a web server, where they found that Copilot could potentially reduce task completion time by a factor of two {\cite{kalliamvakou_2022}}. These and other reports that the code-recommendation systems can improve programmer productivity motivate our research to pursue improvements to these systems.

Code-recommendation systems are powered by large language models (LLMs) such as GPT that are trained on standard language modeling objectives using the Common Crawl data \cite{radford2019language}, and then fine-tuned on public code repositories \cite{chen2021evaluating}. The public roll-out of the code recommendation models has attracted millions of programmers, enabling a unique opportunity to leverage the data of programmers interacting with the models. In this work, we study GitHub's Copilot which is used by millions of programmers \cite{copilot}. For a set of programmers within our organization who consented to have their usage data collected, we collected telemetry data of Copilot suggestions, along  with their associated prompts and the programmer's action to accept or reject the suggestions. We leverage this telemetry data to design mechanisms and interventions that can improve the interaction between programmers and \coderec. 

Specifically, we seek to identify \emph{when} to show a code suggestion. We first define the expected utility of a displaying a suggestion, a value that measures the impact of showing a suggestion on the overall time to write a specific piece of code. This value provides an optimal criterion for when to show a suggestion. However, computing the utility of suggestions is difficult and not currently feasible. Instead, we rely on the result that suggestion utility increases the more likely a suggestion is to be accepted and decreases with increasing latency to generate a suggestion---two quantities we can reliably estimate and control, respectively. We develop a procedure, named {\bf c}onditional suggestion {\bf d}isplay from {\bf h}uman {\bf f}eedback (CDHF) which guides whether to show or hide suggestions. At each pause in keystrokes, CDHF decides whether if it is worthwhile to generate a suggestion and if the programmer is likely to accept the generated suggestion. CDHF employs a cascade of models that predict acceptance of suggestions. The optimization procedure guarantees that any suggestion that was hidden (or not generated) would have been rejected if it was shown with a probability of at least $p$, where, e.g., $p$ can be $0.99$. 

Using data from programming sessions of 535 programmers with feedback on 168k suggestions, we perform a retrospective evaluation of CDHF. We show that we can hide 25\% of suggestions that were shown while guaranteeing that 95\% of them would have been rejected. Further, we avoid generating 13\% of these suggestions. The results show that CDHF would increase the acceptance rate by 7.2\%. The procedure allows for controlling a trade-off that balances the number of suggestions that are displayed with increases in latency, controlled with a parameter that halts generations. We note that a minimal version of CDHF has been implemented in a newer version of GitHub Copilot \cite{newcopilot2023} following the presentation of earlier versions of our work to GitHub. Our paper provides a roadmap for building and fielding better forms of suggestion display. 

Beyond decisions about displaying recommendations, we examine the feasibility of using suggestion acceptance as a reward signal to select \emph{which} suggestions to display and show how partial completions can be prioritized over the generations of complete code segments. While we investigate Copilot in this work, we believe our insights extend to other AI models and non-code-based tasks. Please refer to the arXiv version of this work for an appendix \cite{mozannar2023show}.



\section{Related Work}\label{sec:related_work}
The closest related work to ours is the procedure to selectively hide suggestions in \cite{sun2022learning} (quality estimation before completion, QEBC). In distinction to this work, QEBC \cite{sun2022learning} is not based on human feedback of accepting suggestions but rather is based on constructing a learned estimator of the quality of code completions from datasets of paired code segments and model completions. Our CDHF estimator uses real programmer behavior data and is based on data from a code-recommendation system in current use (Copilot) as opposed to custom-trained ones in \cite{sun2022learning}. 
Different metrics and datasets have been proposed to evaluate the performance of code recommendation models, but these typically assess how well the model can complete code in an offline setting without developer input rather than evaluating how well it assists programmers in situ \cite{ziegler2022productivity,li2022competition,evtikhiev2022out, dakhel2022github}.
Integrating human preferences when training machine learning models has long been studied in the literature \cite{knox2008tamer,macglashan2017interactive}. In particular, reinforcement learning from human feedback (RLHF) has been used to improve LLMs used as conversational chatbots \cite{ziegler2019fine,bai2022training}, notably ChatGPT \cite{chatgpt}. In contrast, CDHF uses human feedback collected organically through telemetry. The objective is fast inference to reduce latency and hiding suggestions rather than updating the LLM. Further related work can be found in the appendix. Our theoretical formulations build on earlier work on harnessing machine learning and utility to guide AI versus human-powered contributions in human-AI interactive settings \cite{horvitz1999principles}, which we apply to our setting. 

\section{Problem Setting}\label{sec:ai_programming}

\textbf{\coderec.} We consider \coderec, which is a commonly used and exemplary tool of AI-powered code recommendations used by millions of programmers \cite{copilot}. \coderec is powered by a large language model (LLM) to provide code suggestions to programmers within an IDE whenever the programmer pauses their typing. An illustration of \coderec suggesting code as an inline, single-colored snippet is displayed in Figure~\ref{fig:ide_coderec}. The programmer can choose to accept this suggestion via a keyboard shortcut (e.g., tab).

\textbf{AI-Assisted Programming.} We attempt a mathematical formalization of programming with the help of a code recommendation model such as \coderec, which we dub \emph{AI-Assisted Programming}. The programmer wishes to complete a certain task $T$, for example, to implement a logistic regression classifier. 
As the programmer writes code starting from time  $0$, \coderec attempts to provide code suggestions at different times. At a given time $t$, \coderec \footnote{We discuss implementation details of Copilot at a high level; our work is based on the August 2022 version of Copilot.} uses a portion of the code $X_t$ to generate a prompt $P_t$, which is passed to the underlying LLM. \coderec then generates a code suggestion $S_t$, which is shown to the user at time $t+\tau$ where $\tau$ accounts for the LLM latency.  Once the suggestion is shown, the programmer must make an action at a certain time $t'>t+\tau$, the action is $A_{t'} \in \{\textrm{ accept, reject}\}$; the \emph{reject} action is triggered implicitly by continuing to type.  

\begin{figure}[t]
	\centering
\includegraphics[width=0.6\textwidth]{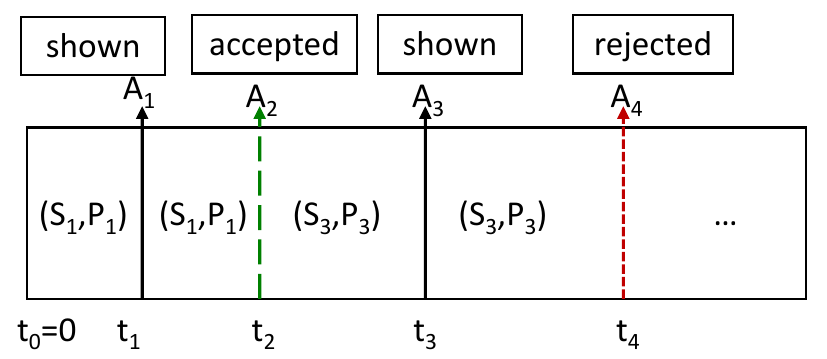}
	\caption{Schematic of telemetry with \coderec as a timeline. For a given coding session, the telemetry contains a sequence of timestamps and actions with associated prompts and suggestions.}
	\label{fig:telemetry}
\end{figure}

\textbf{Telemetry.} \coderec logs aspects of the interactions via telemetry, which we leverage in our study. We refer to event positions drawn from a discretization of times spanning a session. 
Specifically, whenever a suggestion is shown, accepted or rejected, we record a tuple to the telemetry database, $(t_i, A_i, P_i, S_i)$, where $t_i$ represents the within-session timestamp of the $i^\text{th}$ event ($t_0 = 0$), $A_i$ details the action taken (augmented to include `shown'), and $P_i$ and $S_i$ capture the prompt and suggestion, respectively. Figure~\ref{fig:telemetry} displays a portion of timeline built from telemetry data drawn from a coding session.  Telemetry data from each programmer is stored in a database $D=\{(t_i, A_i, P_i, S_i)\}_{i=1}^n$  and represents a discretized representation of the interaction and provides the human feedback data we leverage.

\textbf{Programmer State.} When faced with a suggestion, is a programmer looking and verifying it, or rather engaged in other activities such as thinking about their code or looking at documentation? The state of the programmer is important in the expected value of the recommendation. However, we cannot answer this question as the telemetry does not capture the programmer's activities and thinking between two consecutive time stamps $t_i$ and $t_{i+1}$, i.e., the space in between the arrows in Figure~\ref{fig:telemetry} which we refer to as the \emph{programmer's latent state}. In an earlier publication \cite{mozannar2022reading}, we describe a study of 21 participants focused on gaining an understanding of sequences of states visited during the writing of code, including latent states. The work employed videos and interviews to acquire information about the latent states. We showed in the work that including information about latent states can significantly boost predictions about accepting recommendations, motivating the collection of data beyond that captured in telemetry. 

In this work, we endeavor to understand the impact of the programmer latent state denoted as $\phi_t$ and its effect on our ability to leverage the telemetry (human feedback data) to improve AI-code recommendation systems. 


\section{Theoretical Formulation of Suggestion Utility}\label{sec:utility_theory}


A critical design question in programmer-{\coderec} interaction is \textbf{when} should the model inject a suggestion into the IDE? 
The version of that we {\coderec} provides a suggestion when it detects a brief pause in the IDE. Alternative interaction designs would require the programmer to ask for suggestions using a keyboard shortcut or to enable a mix of human and machine initiatives. Requiring the programmer to ask may lead to sub-optimal interactions because its success would rely on programmers having an accurate mental model of {\coderec} abilities {\cite{sarkar2022like}} which can require long-term interactions with the model \cite{bansal2019beyond} or training \cite{mozannar2022teaching}. Second, requiring an
explicit invocation can disrupt the natural flow of programming, breaking a state of flow achieved during intensive focus {\cite{csikszentmihalyi2014flow}}. Designs requiring user initiative as well as those automatically displaying content can burden users with interruptions that decrease task performance \cite{bailey2001effects,Cutrell2001Interrupt}. We note that such costs can be inferred and accounted for formally in utility-theoretic systems \cite{HA_Interrupt2003,Attention_sensitive1999}. 

Ideally, {\coderec} should display suggestions when the suggestions provide net value to programmers. For example, consider the task of completing a function and the time taken to complete it as a proxy for the total effort. If the expected time required to verify and edit {\coderec}'s suggestion exceeds the time to write the code by themselves (counterfactual cost), then {\coderec} should not show its suggestions. Conversely, if the expected time to write exceeds the time to verify and edit, it may be useful to display the suggestion. We now formalize this intuition with a utility-theoretic formulation and, in the next section, discuss the methodology to make it practical.  

\textbf{Programmer Model.} At a given time instance time during a session, \coderec extracts a prompt $P$ from the code file $X$ and generates a code suggestion $S$. If this suggestion is shown, we assume the programmer spends an expected time $\E[\mathrm{verification}|X,S,\phi]$ to verify it and accepts the suggestion with probability $\bP(A=\textrm{accept}|X,S,\phi)$.
Once a suggestion is accepted, the programmer may further edit the suggestion with expected time $\E[\mathrm{editing}|X,S,\phi,A=\mathrm{accept}]$ to achieve their task. On the other hand, if the programmer rejects the suggestion, they would have to spend time writing code that achieves their task, denoted by $\E[\mathrm{writing}|X,S, A=\mathrm{reject}]$. 
Thus, the total time incurred with showing a suggestion, denoted as $\E[\mathrm{S \ shown}|X,S,\phi]$, is:
\begin{align}\label{eq:time_shown}
	&\E[\mathrm{S \ shown}|X,\phi] = \E[\mathrm{verification}|X,S,\phi] 
	\\&+ \bP(A=\textrm{accept}|X,S,\phi) \cdot  \E[\mathrm{editing}|X,S,\phi,A=\mathrm{accept}]  \nonumber\\
	&+\bP(A=\textrm{reject}|X,S,\phi) \cdot  \E[\mathrm{writing}|X,S, \phi, A=\mathrm{reject}] \nonumber
\end{align}

While editing and writing, \coderec may further make more suggestions; thus, the editing time and writing time should include interactions with future suggestions. 
Now, on the other hand, if the suggestion is \emph{not} shown,  the programmer will spend time $\E[\mathrm{writing}|X]$ writing code for their task.
We also need to factor in latency, the time cost $\tau$ to compute a suggestion once we decide to create a suggestion. Latency is only experienced by the programmer if their latent state $\phi$ includes expecting a suggestion and waiting for it. If the programmer is expecting a suggestion, we should add $\tau$ to the total time when we show a suggestion; otherwise, the programmer continues to write code not expecting a suggestion. 

We now define suggestion utility, a value that indicates the change in programmers' coding time due to showing the suggestion.
\begin{definition}[Suggestion Utility]
	The time impact $\delta$, denoted as the \emph{suggestion utility}, of showing $S$ versus not showing is defined as:
	\begin{align}\label{eq:delta}
		&\delta =   
		\underbrace{\E[\mathrm{writing}|X,\phi]}_{\mathrm{S \ not \  shown}} - \underbrace{\E[\mathrm{S \ shown}|X,\phi]}_{\mathrm{S\ shown}}  - \underbrace{\E[\tau|X,\phi]}_{\mathrm{latency}}
	\end{align}
\end{definition}
From the above, a suggestion $S$ at a given time should \textbf{only be shown} if $\delta >0$ (Equation~\ref{eq:delta}), where the programmer will spend less time to achieve their task if it is shown.
An optimal scheme to know when to show suggestions would be to generate suggestions as frequently as possible, compute their \emph{suggestion utility} $\delta$, and display them if $\delta >0$.

\textbf{Feasibility of Estimating $\delta$.} Per Equation \eqref{eq:time_shown}, computing \emph{suggestion utility} requires the computation of four quantities: (1) the expected time spent verifying a suggestion, (2) the expected time editing a suggestion, (3) the expected time to write a segment of code and (4) the probability of accepting a suggestion. One can attempt to build an estimator for (1), by predicting from the prompt and suggestion the time spent verifying a suggestion $i$ which would be $t_{i+1}-t_{i}$ using standard regression estimators. Unfortunately, using the same features and the dataset detailed in our experimental section, our best estimator is only able to achieve an $R^2=0.13$, which is not much better than a naive median time estimate. This may be due to the high variance and unobserved confounders governing verification time. Estimating editing and verification time (quantities 2 and 3 above) is only more complex and challenging. Thus, we restrict our methodology to seeing when we can evaluate $\delta$ using only our estimator for the probability of acceptance (4). 

\textbf{Learning Programmer's Acceptance Decisions.} The full conditional for the probability that the programmer accepts a suggestion is $\bP(A=\textrm{accept}|X,S,\phi)$. Given the telemetry, we can only compute $\bP(A=\textrm{accept}|X,S)$ where the programmer's latent state cannot be observed. Using standard calibrated classification methods, we can estimate the probability $\bP(A=\textrm{accept}|X,S)$ by using the actions $A_i$ as the labels. 
We show that a simple mechanism of thresholding the estimated probability that the programmer accepts a suggestion is equivalent under certain assumptions to checking if $\delta < 0$:
\begin{proposition}\label{prop:pstar}
Under assumptions that the programmer spends more time writing code when they reject a suggestion compared to when they accept a suggestion and edit it, given specific code, suggestion, and latent state $(X,S,\phi)$, if the programmer's probability of accepting $\bP(A=\textrm{accept}|X,S,\phi)$ a suggestion is below $\bP^*$, which is defined as:
	\begin{align}\label{eq:p_star}
		& \bP^* = \frac{\E[\mathrm{verification}] + \E[\mathrm{latency}] }{   \E[\mathrm{writing}|A=\mathrm{reject}] - \E[\mathrm{editing}|A=\mathrm{accept}] } 
	\end{align}
	then the suggestion should not be shown. Note that $\bP^*$ is defined as a function $\bP^*(X,S,\phi)$ evaluated pointwise. 
\end{proposition}

\begin{figure}[t]
	\centering
	\includegraphics[width=0.7\textwidth]{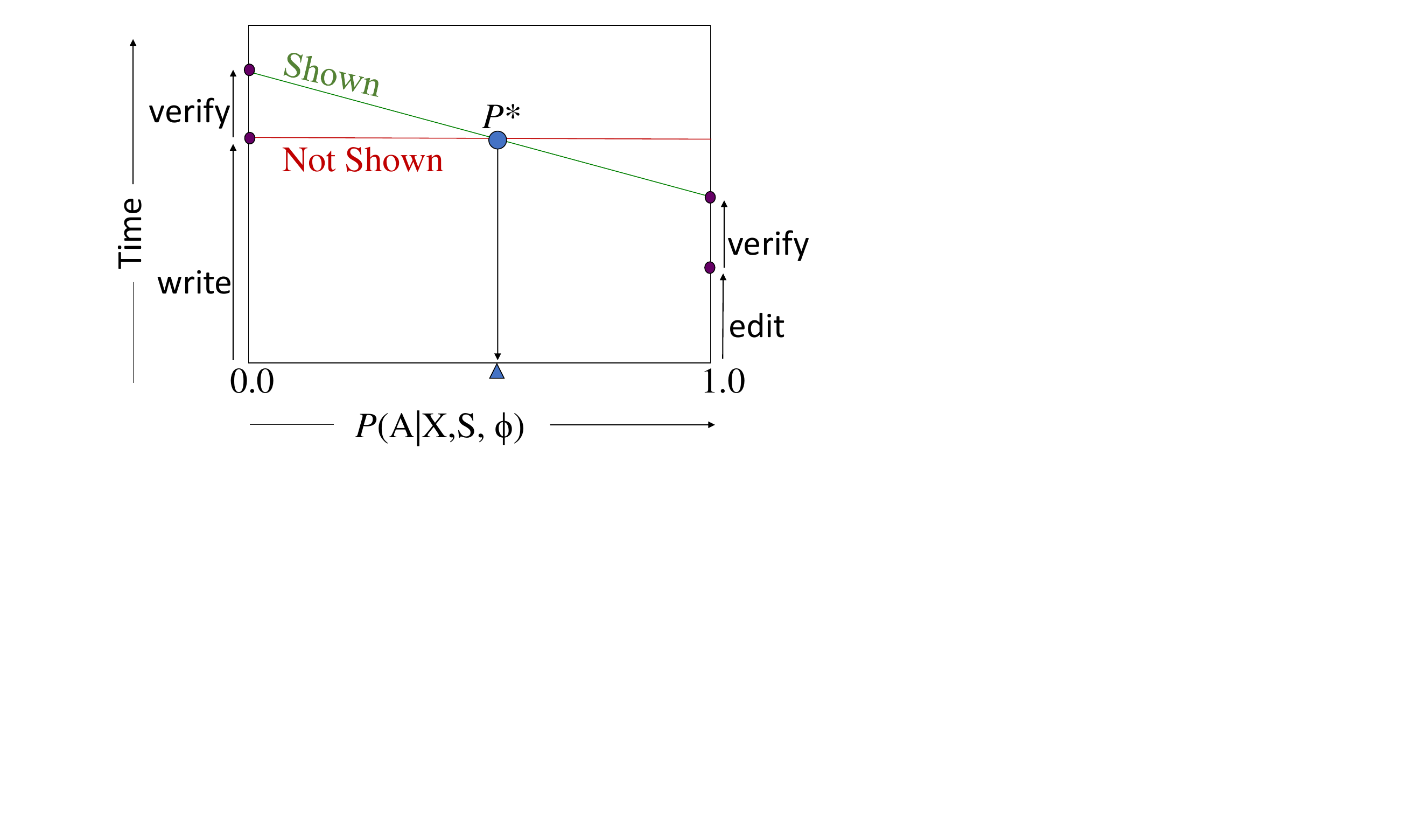}
	\caption{Graphical depiction of analysis of Proposition \ref{prop:pstar} when the latency is zero. The y-axis shows total time and the x-axis is the programmer's probability of accepting $\bP(A=\textrm{accept}|X,S,\phi)$. At probability $\bP^*$, showing and not showing the suggestion have equal time cost.   }
	\label{fig:pstar}
\end{figure}

The formal statement and proof are available in the appendix. The above proposition shows that comparing the probability of acceptance to $\bP^*$   can guide when to show the suggestion. We provide a graphical view of the analysis in Figure \ref{fig:pstar}, in the spirit of related analyses on utility-guided interactive interfaces \cite{horvitz1999principles}. Practically, if we compare the probability of acceptance to a constant lower bound of $\bP^*$, we can guarantee that we hide suggestions only when $\delta <0$.

\textbf{Effect of Programmer Latent State.} As mentioned previously, the programmer's latent state is not available via telemetry. Thus, we can only provide predictions of $\bP(A=\textrm{accept}|X,S)$ versus explicit consideration of the latent state, $\bP(A=\textrm{accept}|X,S,\phi)$. In earlier work \cite{mozannar2022reading}, we collected telemetry data of 21 programmers performing various tasks and had participants retrospectively label the telemetry with their latent state from a set of twelve unique states (1096 suggestions). We use this data to build predictive models with and without the latent state using the same methodology in the experiments section. 
Using a leave-one-out programmer evaluation strategy, the model without the latent state achieves accuracy $61.9 \pm 1.9$ while the model with the latent state achieves $83.6 \pm 2.4 $, a statistically significant difference according to a paired t-test ($p=6.9e-7,t=7.11$); a similar result occurs when comparing areas under the receiver operating characteristic curve (AUC). These results highlights an opportunity to gather external data beyond telemetry to build such predictive models and indicates that acceptance may not simply be a property of suggestions and code context.

\section{Conditional Suggestion Display From Human Feedback}\label{sec:cdhf}

In this section, we describe the CDHF method that can be implemented using telemetry data to identify when to show suggestions, as illustrated in Figure~\ref{fig:ide_coderec}. We note from Equation \eqref{eq:p_star} that the higher the probability of accepting the suggestion and the lower the latency to generate the suggestion, the more likely the suggestion is useful ($\delta>0$). Our proposed approach is as follows: Each time the programmer pauses typing, we decide using a predictor whether to show a suggestion. Crucially, we do this using a two-stage scheme to avoid generating suggestions when we know the programmer would reject them.

\textbf{Display Decision.} Let $m(X,S)$ be a binary predictor that denotes whether, at a given moment in the code $X$, we should show the suggestion $S$; we call this the display decision. If $m(X,S)=1$, we display the suggestion; otherwise, we do not.  
The most straightforward way to build such a function $m$ is to estimate the programmer's probability of accepting the suggestion: $\bP(A=\textrm{accept})|X,S)$ and then threshold the probability so that suggestions that fall below a probability $t$ are hidden. However, this will lead us to generate suggestions including those that will never be shown, thus wasting computing resources. We propose to decompose the function $m$ so that we first decide using only the code whether we can make the display decision without generating the suggestion $S$ with a function $r(X)$. If $r(X)=1$, we make the display decision using a stage 1 model $m_1(X)$ without generating the suggestion, otherwise if $r(X)=0$ we generate the suggestion $S$ and make the display decision with a stage $2$ model $m_2(X,S)$  as follows: 
\begin{equation}
m(X,S) = r(X) \cdot  m_1(X) + (1-r(X)) \cdot m_2(X,S)
\end{equation}
This formulation allows us to avoid generating suggestions when we can make an accurate display decision in advance of knowing the suggestion. For example, in a setting where the programmer has rejected the last 30 suggestions, they are unlikely to accept the next suggestion. 

\textbf{Objective and Guarantees.} Our objective in learning the functions $r,m_1,m_2$ is to (1) hide as many suggestions that would have been rejected and (2) maximize the number of display decisions made without generating the suggestion to reduce latency on the system. There is an inherent trade-off between these two objectives as making decisions with access to the suggestions would be more accurate. Moreover, we want to make sure we do not hide suggestions that would have been accepted, as this would limit the usefulness of the code assistant. Therefore, we impose a constraint that, whenever we hide a suggestion, there is at least a probability $p$ it would have been rejected, a constraint on the true negative rate (TNR). We translate the objectives and the constraint into the following optimization problem:
\begin{align} \label{eq:cdhf_opt_obj}
& \max_{r,m_1,m_2} \lambda \bE[1 - m(X,S)]  + (1-\lambda) \bE[r(x)] \\ 
& s.t. \ \ \bP(A=\textrm{reject}|m(X,S)=0) \geq p
\end{align}

\textbf{Parameterization.} We can control the trade-off between the two objectives with a hyperparameter $\lambda \in [0,1]$. Equivalently, instead of controlling the trade-off with $\lambda$, we can set a constraint on $\bE[r(x)]:=R$ and set $\lambda=1$.  We propose an intuitive post-hoc procedure to solve the optimization problem \eqref{eq:cdhf_opt_obj}: We first learn calibrated estimators of the probability of accepting suggestions $\hat{\bP}(A=\textrm{accept}|X,S)$ (with suggestion) and  $\hat{\bP}(A=\textrm{accept}|X)$ (without suggestion). We then parameterize: $$m_1(X)=\bI_{\hat{\bP}(A=\textrm{accept}|X) \geq t_1 }, m_2(X)=\bI_{\hat{\bP}(A=\textrm{accept}|X,S) \geq t_2 }$$ and $r(X) = \bI_{H(\hat{\bP}(A=\textrm{accept}|X)) \leq t_r }$ ($H(.)$ is Shannon's Entropy), and optimize \emph{jointly} over the tuple of thresholds $t_1,t_2,t_r$ over $[0,1]^3$.  This is a fairly efficient procedure that can achieve good results. We note that this procedure saves latency indirectly by reducing the number of LLM calls across the session and across different users, and that we should still enable the user to see the suggestion with a special keyboard shortcut to override the display decisions. In the next section, we perform a retrospective evaluation of CDHF.

\begin{figure*}
	\centering
	\includegraphics[width=1.0\textwidth]{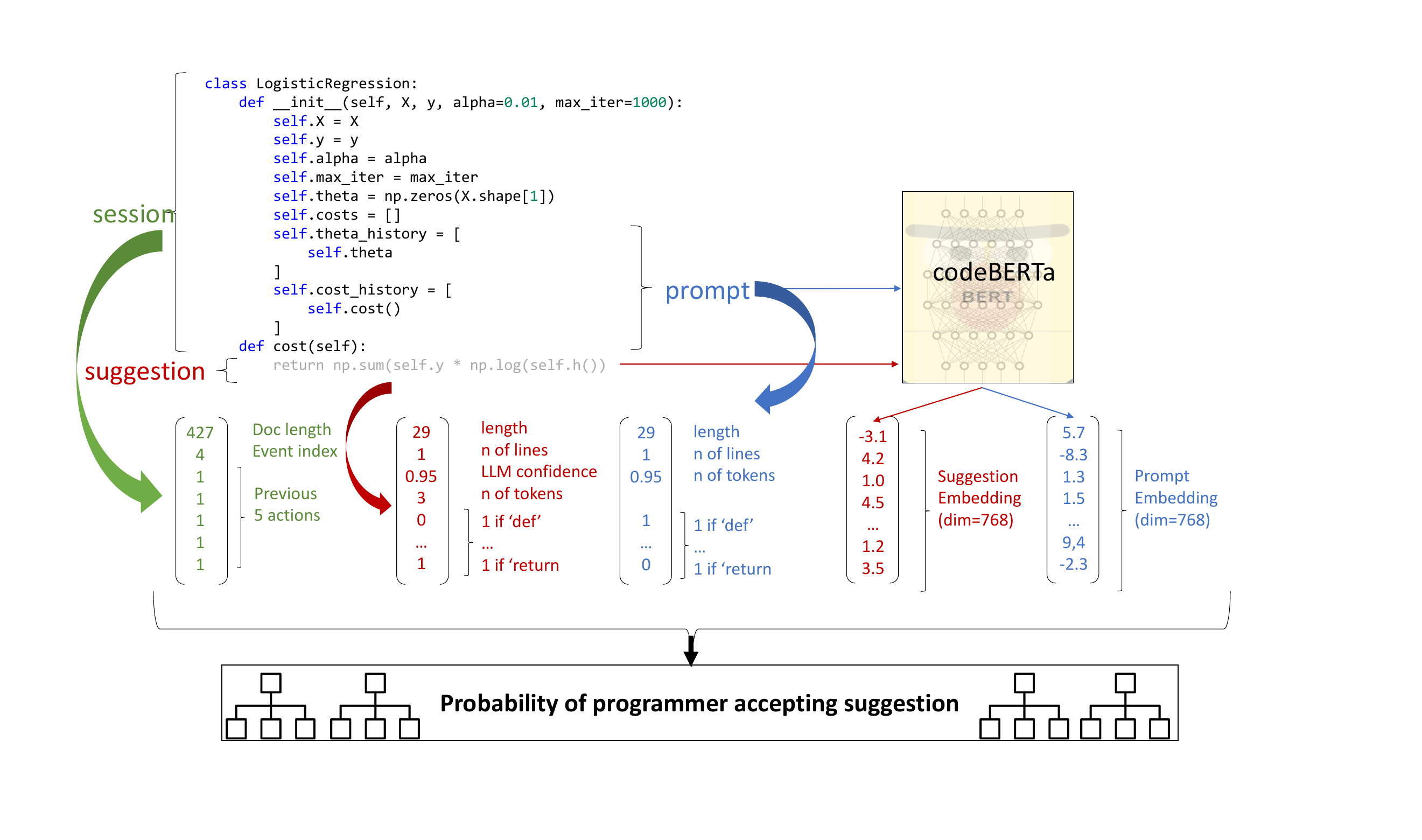}
	\caption{Features used to build action prediction model in  Experiments \ref{sec:experiments}, including from the suggestion, prompt, and session.}
	\label{fig:model_predict}
\end{figure*}

\section{Experiments}\label{sec:experiments}
Our main aim with experiments is to understand how well the CDHF procedure can make display decisions in a retrospective evaluation. Code is available\footnote{\url{https://github.com/microsoft/coderec_programming_states}} and additional details can be found in the appendix.

\subsection{Dataset and Feature Engineering.}

\textbf{Dataset.} To build and evaluate our methods, we extract a large number of telemetry logs from \coderec users (mostly software engineers and researchers) at Microsoft. Programmers provided consent for the use of their data, and its use was approved by Microsoft's ethics advisory board. Specifically, for a two-week time period, we extracted all the telemetry events for 535 users who coded in Python. This totals 4,749 coding sessions, where a session is defined as a continuous sequence of user actions with at most 30 minutes between consecutive events. These sessions are from real-world usage of {\coderec} for daily tasks of the software engineers and researchers, the data was collected prior to the inception of our work. On average, each user contributes nine sessions, with each session lasting 21 minutes  (median, 12 minutes). Sessions contain an average of 97 events (show, accept, and reject). This totals to almost 1,675 hours of coding with 168,807 shown events and 33,523 accept events, yielding an acceptance rate of 21.4\% (not meant to represent Copilot's average acceptance rate).

\textbf{Model Features.} The telemetry dataset $D$ described above contains for each user a list of events in each of their coding sessions; we denote $D_{i,j}$ to be the list of events for the j'th session of the i'th user.  The dataset $\mathbf{D} = \{ D_{i,j} \}$  contains for each user $i$ and session $j$, a list of events occurring in the corresponding coding session. We extract only the accept and reject events, as well as prompt and session features of the corresponding shown events. For each prompt and suggestion pair, we extract: the programmer id as one hot vector,
the length of the document, the previous five actions, suggestion features (e.g., suggestion length), previous features of the last five suggestions shown, the confidence reported by \coderec, an embedding using codeBERTa \cite{feng2020codebert} of prompt and suggestion, presence of Python keywords (e.g., \texttt{import}, \texttt{def}  \texttt{try}, etc.), and the output of the Tree-sitter Parser~\cite{tree_sitter}.
Finally, we extract features of the prompt, including its embedding, textual features, and parser outputs.
Figure \ref{fig:model_predict} summarizes the feature engineering. It is crucial to note that the features do not leak any information about future events and can be computed as soon as a suggestion is generated by {\coderec}. For the first stage model ($m_1$) in CDHF, suggestion features are omitted while we include all features for the second stage model ($m_2$). This feature engineering incorporates past actions and suggestions that the programmer has seen and allows us to use regular ML algorithms instead of time-series methods. 

\subsection{Model Evaluation}

Before we evaluate CDHF, we perform an evaluation of the programmer acceptance model $m_2(X,S)$. We split the telemetry dataset in a 70:10:20 split for training, validation, and testing respectively.  Importantly, we do this split in two ways: (1) by randomly splitting over programmers so that no single programmer is shared across the three splits and, (2) by randomly splitting over sessions so that users in training can also be seen in testing to allow for personalization.

\textbf{Results.} We evaluate different standard machine learning models on this task and find that the best-performing model is eXtreme Gradient Boosting (XGB) \cite{chen2015xgboost}. When we split across users, XGB is able to achieve 81.1\% (95\% CI 80.7-81.6 ) accuracy and, more importantly, 0.780 (95\% CI 0.775-0.786) AUC. In the appendix, we show metrics for different models evaluated, including deep networks \cite{mozannar2023show}.
The results indicate that the model is able to distinguish between suggestions that are likely to be accepted versus those likely to be rejected. The model is also well calibrated: the expected calibration error is 0.10 \cite{naeini2015obtaining}. 

We note a significant increase in AUC when we allow for personalization: including programmer ID as a feature and splitting across sessions, this leads to an AUC of 0.795 (95\% CI 0.789-0.801), a significant increase (basis of $m_2$ model). When we remove suggestion features from the model, the resulting model (basis of $m_1$ model) achieves an AUC of 0.631 (95\% CI 0.624-0.638). The time to compute the features needed for the models and performing inference on a single data point can take 10ms with a GPU and less than 1ms on a CPU when omitting embeddings, in addition to latency of sending and receiving information between server and client. 
In the appendix, we show results for different ablation of model features, sample complexity plots, and feature importance plots. 

\subsection{Retrospective Evaluation of CDHF}

We train the models $m_1$ and $m_2$ using the training set per the previous subsection. We set the thresholds $t_1,t_2,t_r$ on the validation set for CDHF and evaluate on the test set. 

\begin{figure}
	\centering
	\includegraphics[width=0.7\textwidth]{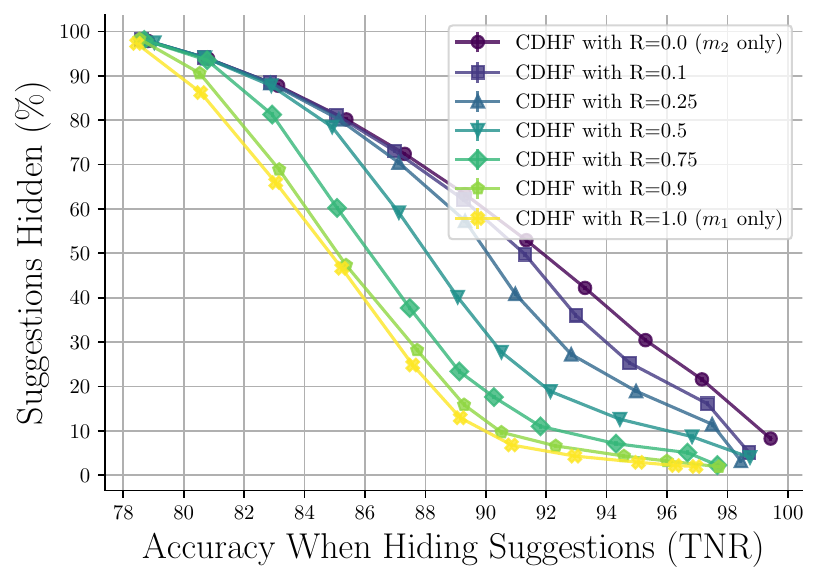}
	\caption{Evaluation of CDHF for selectively hiding suggestions. For a given constraint on FNR (accuracy when a suggestion is hidden) on the x-axis, we show on the y-axis the fraction of the total suggestions we can hide while guaranteeing the desired FNR. We plot these curves while varying how often the decision is made generating suggestions (R:=$\bE[r(x)]$, when R=0, we generate the suggestion then decide to hide or not, when R=1, we decide to hide without knowing the suggestion).}
	\label{fig:cdhf_results}
\end{figure}

\textbf{Results.} In Figure \ref{fig:cdhf_results}, we vary the desired TNR rate (accuracy when a suggestion is hidden) and plot how many suggestions we can hide from those previously displayed while guaranteeing the desired TNR rate. We show the behavior of the CDHF method with different $\lambda$ values, or, equivalently, with different constraints on how often the $m_1$ model (first stage) is used: $R:=\bE[r(x)]$. To illustrate what CDHF can accomplish, we can hide 25.3\% of suggestions that were shown while guaranteeing that 94.7\% of them would have been rejected and avoid generating 12.9\% of the suggestions. If we have no concerns for latency, we can hide 52.9\% of suggestions while guaranteeing that 91.3\% of them would have been rejected. Figure \ref{fig:cdhf_results} shows how we can achieve different trade-offs by selecting an operating point on any given curve.
CDHF is able to satisfy the constraint of FNR on the test set with a violation of at most 0.3\% i.e., a guarantee of 95\% FNR on the validation set equates to 94.7-95.3\% on the test set.

\textbf{Counterfactual Increase in Acceptance Rate.} On the test set, the acceptance rate of suggestions is 22.5\%. Retrospectively, if we had used CDHF to hide 52.9\% of suggestions, we could compute a counterfactual acceptance rate. The counterfactual acceptance rate can be computed as: $\frac{\textrm{S accepted} \cdot (1-\textrm{\% hidden} \cdot (1-TNR) )  }{\textrm{S shown} \cdot \textrm{\% not hidden}} = \frac{ 22.5 ( 1-0.529 \cdot 0.087)}{0.471}=45.6\%$, which is a  23.1 point increase, a value we expect to be an overestimate.

\textbf{Discussion and Limitations.} The retrospective evaluation shows that CDHF has promise in reducing developer time spent verifying suggestions or waiting for suggestions. We note that our evaluation is retrospective. Although GitHub has shown that conditional suggestion filters similar to CDHF increase the acceptance rate of suggestions, a user study is required to verify whether the method makes programmers more productive. As Goodhart's law states, once a metric becomes a target, it ceases to become a good measure; acceptance rate is no exception. Moreover, if CDHF is not trained with sufficient data that captures the programmer's use cases, it can make the programming experience worse by hiding useful suggestions.  Moreover, a rejected suggestion may still be useful, which we do not account for here. Finally, the optimization problem in \eqref{eq:cdhf_opt_obj} is amenable to procedures inspired by learning to defer \cite{mozannar2020consistent} that can outperform the post-hoc procedure proposed.

\section{Which Suggestion to Show?}

We focus in this study on the problem of when to display suggestions. We did not tackle the question of which suggestions to display among a candidate set. Given access to telemetry data, which consists of contextualized suggestions with accept and reject signals, one can interpret an accept as the act of preferring a suggestion over no suggestion. It is reasonable to harness the telemetry data as a preference dataset and build a reward model of programmers' preferences, which would be equivalent to estimating the programmer's acceptance probability $\hat{\bP}(A=\textrm{accept}|X,S)$. Thus, a reasonable procedure is to take a candidate set of suggestions $\mathcal{S}$ and display the suggestion that maximizes the probability of acceptance across the set; this is essentially the best-of-$n$ baseline approach in RLHF \cite{rafailov2023direct}.

\textbf{Potential Bias Towards Short Suggestions.} We hypothesize that such a ranking scheme would not be productive and can lead to poor suggestions of short length. Our rationale is the following: suppose the LLM is able to generate a multi-line suggestion $S$ for a user query that approximately matches what the user desires. To maximize the probability that the user accepts the suggestion, it would be advantageous to split the suggestion $S$ line-by-line and display it to the user step-by-step. The reasoning is that it is more likely for the first line of $S$ to be correct rather than all of $S$ being correct, hence being more likely to be accepted.

\textbf{Experiment.} To test this hypothesis, we perform the following experiment: We learn a model $m$ of suggestion acceptance given only the prompt and suggestion embeddings with no session features on the telemetry data from the previous section. We then leverage the HumanEval dataset \cite{chen2021evaluating}, which consists of 164 Python problems, each with an associated docstring and a ground truth function body solution. Solutions have at least two lines and seven median lines of code. Given the model $m$ and each problem, we let the prompt be the concatenation of the docstring and the first $k$ lines of the solution and let the candidate set of suggestions $\mathcal{S}$ be as follows: Given the solution $S$ represented as an array of tokens of length $N$, we let $\mathcal{S}=\{S[:i]\}_{i=1}^{N}$. For example, if the solution $S$ was "return np.mean(x)", then $\mathcal{S}=\{\textrm{"return"},\textrm{"return np.mean(x)"} \}$.

\textbf{Results.} We vary the parameter $k$ in the set $\{0,1,2,3\}$ so that the prompt goes from the docstring to include lines of the solution. When $k=0$, the normalized length of the highest-rated suggestion, according to the model across the 164 problems, is almost uniform across $[0,1]$, a Kolmogorov–Smirnov test compared to the uniform distribution has a p-value of 0.53 (KS=0.06).  
Optimally, we want the normalized length to cluster around $1$ to include the full solution. However, when $k>0$, meaning that the prompt includes lines of code, we find that for over 60 of the 164 problems, the highest-scored suggestion lies in $[0,0.2]$, and, for at least 40 problems, it is the first token. A histogram showing this phenomenon is in Figure \ref{fig:which_k1} for $k=1$.  This provides some evidence that optimizing for acceptance can be biased toward shorter suggestions since the highest-ranked suggestion is often in the first few lines of the solution.

\textbf{Limitations.} However, there are important limitations in our experiment. First, the model $m$ is only trained on Copilot suggestions. Thus, the bias towards short suggestions can be due in part to Copilot potentially showing short suggestions. Maximizing acceptance would not alleviate such a bias. Second, while the use of embeddings of the suggestions for the reward model led to accurate predictions of accepts (AUC=0.701), they might be biased in some ways as compared to alliance on fine-tuning the language model.

\begin{figure}[H]
  \centering
  \subfigure[Histogram of position of max suggestion]{
    \includegraphics[width=0.45\textwidth]{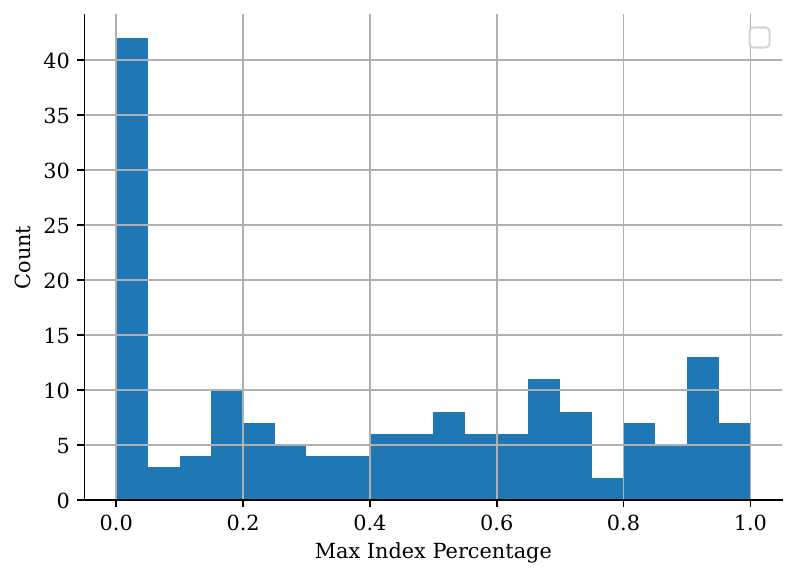}
    \label{fig:plot1}
  }
  \hfill
  \subfigure[Probability of acceptance by length]{
    \includegraphics[width=0.45\textwidth]{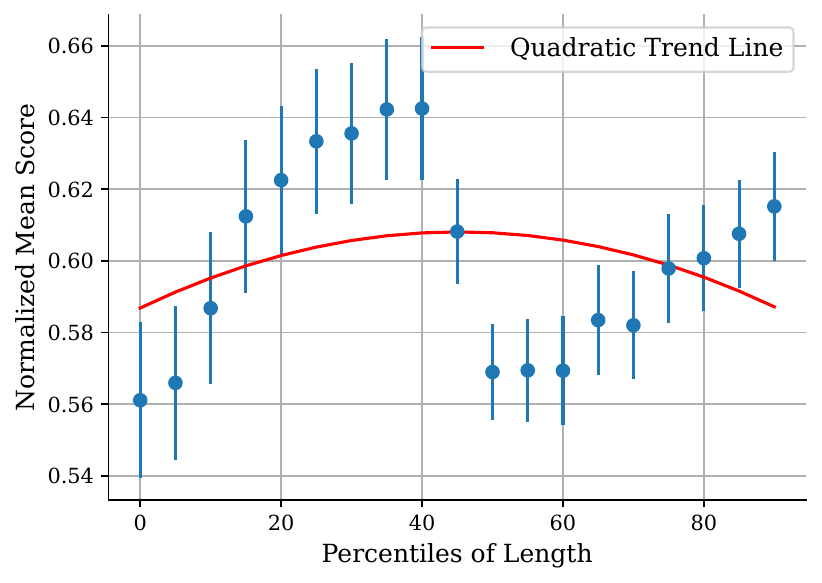}
    \label{fig:plot2}
  }
  \caption{Plots for the experiment on ranking suggestions by the probability of acceptance. Histogram (a) shows in which length percentile bin the suggestion maximizing the probability of acceptance lies and Graph (b) shows the acceptance score by increasing the length of the suggestion. These plots are for $k=1$ (docstring + first line of solution) }
  \label{fig:which_k1}
\end{figure}

\section{Conclusion}\label{sec:conclusion}

We proposed a strategy to decide when to display code suggestions in AI-assisted programming to improve time efficiency. This strategy was based on a utility theory formulation and employs a two-stage procedure using a predictive model of suggestion acceptance. A retrospective evaluation showed that we can reduce the number of suggestions and thus programmers' time without sacrificing the utility of \coderec. However, a prospective study that evaluates the impact of \coderec with and without CDHF could help with conclusive evaluation and is the basis of future work. Moreover, future work will attempt to directly estimate Proposition 1 leveraging improved methods. We don't believe that CDHF can introduce negative consequences beyond what \coderec introduces to the programmer as it functions as a filtering mechanism for unhelpful suggestions. Moreover, we believe that the CDHF methodology can be employed in a wide range of streaming human-AI collaboration tasks such as assisted writing. Future work will incorporate the latent state of the programmer into the predictive models, investigate how to rank suggestions using the models from CDHF, and validate the efficacy of CDHF in user studies.

\section*{Acknowledgments}

Hussein Mozannar partly conducted this work during an internship at Microsoft Research (MSR). We acknowledge valuable feedback from colleagues across MSR and GitHub including Saleema Amershi, Victor Dibia, Forough Poursabzi, Andrew Rice, Eirini Kalliamvakou, and Edward Aftandilian.

\bibliographystyle{alpha}

\bibliography{ref}
\clearpage
\appendix

\section{Extended Related Work}

\paragraph{AI-Assisted Programming.} Large language models (LLMs) such as GPT-3 \cite{brown2020language}, have been widely used in natural language processing. One example of this is Codex \cite{chen2021evaluating}, a GPT model trained on 54 million GitHub repositories, which demonstrates the effectiveness of LLMs in solving various programming tasks. For instance, Codex was tested on the HumanEval dataset of 164 programming problems, where it was asked to write the function body from a docstring and achieved 37.7\% accuracy with a single generation \cite{chen2021evaluating}. Different metrics and datasets have been proposed to evaluate the performance of code recommendation models, but these typically assess how well the model can complete code in an offline setting without developer input, rather than evaluating how well it assists programmers in-situ \cite{hendrycks2021measuring,li2022competition,evtikhiev2022out, dakhel2022github}.
Researchers have found that developers do not need a perfect recommendation model for it to be useful. Weisz et al. conducted interviews with developers and found that they did not require a perfect recommendation model for the model to be useful \cite{weisz2021perfection}, while Ziegler et al. surveyed over 2,000 Copilot users and found that they felt more productive using Copilot \cite{ziegler2022productivity}. A study by Google  found that an internal \coderec model had a 6\% reduction in 'coding iteration time' \cite{google_ai_blog_2022}. On the other hand, a study of 24 participants by Vaithilingam et al. showed no significant improvement in task completion time, yet participants stated a clear preference for Copilot \cite{vaithilingam2022expectation}. \cite{barke2022grounded} showed that interaction with Copilot falls into two broad categories: the programmer is either in 'acceleration mode' or in 'exploration mode'. A similar strategy to selectively hide code suggestions was proposed in \cite{sun2022learning} (Quality Estimation Before Completion, QEBC). However, QEBC \cite{sun2022learning} is not based on human feedback of accepting suggestions, but rather is based on constructing a learned estimator of the quality of code completions from datasets of paired code segments and model completions. In contrast, our CDHF estimator uses real programmer behavior data and is based on data from a code-recommendation system in current use (Copilot) as opposed to custom-trained ones in \cite{sun2022learning}.

\textbf{Human Feedback.} Integrating human preference when training machine learning based models have long been studied in the literature \cite{knox2008tamer,macglashan2017interactive}. In particular, Reinforcement Learning from Human Feedback is an approach where the designer first gathers explicit human preference over actions which is used to improve the model using RL \cite{christiano2017deep}. More recently, this approach has been used to improve LLMs for different tasks \cite{ziegler2019fine,bai2022training} notably including ChatGPT \cite{chatgpt} and consists of three steps: gather human preference over options, train a reward model of human preference, use the reward model to update the LLM using RL. In contrast, our approach CDHF uses implicit human feedback through telemetry more readily collected which is not fully reflective of true preference and avoids updating the LLM. Collaborative filtering employs  human preference data to re-rank content  \cite{su2009survey}, though avoids the complexities of both generating and ranking content which is required here. Our theoretical formulations in Section~\ref{sec:utility_theory} build on the work of interactive user interfaces \cite{horvitz1999principles,dudley2018review} and generalize them to our setting. Work on algorithmic deferral \cite{madras2018predict,mozannar2020consistent,okati2021differentiable} investigates a similar question on whether the human or the AI should accomplish the task and is based on error estimation (instead of time cost estimation here) of the human versus the AI, as well as work on  personalized decision support policies
\cite{bhatt2023learning} investigates whether support should be shown but not when.

\section{Derivation of $\bP^*$}\label{apx:proof_p*}

\textbf{Proposition 1.} \textit{ \noindent Under assumptions of the programmers model, notably that the programmer spends more time writing code when they reject a suggestion compared to when they accept a suggestion and edit it. Given  specific code, suggestion and latent state $(X,S,\phi)$, if the programmer's probability of accepting $\bP(A=\textrm{accept}|X,S,\phi)$ a suggestion is below $\bP^*$ defined as:
\begin{align*}
    & \bP^* = \frac{\E[\mathrm{verification}] +\E[\tau|X,\phi] }{   \E[\mathrm{writing}|A=\mathrm{reject}] - \E[\mathrm{editing}|A=\mathrm{accept}] } 
\end{align*}
then the suggestion should not be shown. Note that $\bP^*$ is defined as a function $\bP^*(X,S,\phi)$ evaluated pointwise. 
}

\begin{proof}

Starting from the equation $\delta =0$, assume that the programmer has only two actions of accept or reject. This is backed by our analysis of our telemetry sample where we noticed that less than 1\% of suggestions are browsed.  
We first have: 
\begin{align}
    & - ( \E[\mathrm{verification}|X,S,\phi] 
   + \bP(A=\textrm{accept}|X,S,\phi) \cdot  \E[\mathrm{editing}|X,S,\phi,A=\mathrm{accept}]  \nonumber\\
    & + \bP(A=\textrm{reject}|X,S,\phi) \cdot  \E[\mathrm{writing}|X,S, \phi, A=\mathrm{reject}] ) + \E[\mathrm{writing}|X,\phi] -\E[\tau|X,\phi]  =  0 \nonumber
\end{align}
We now replace $\bP(A=\textrm{accept}|X,S,\phi)$ by $\bP^*(X,S,\phi)$ and move around terms:

\begin{align*}
    & \bP^*(X,S,\phi) \cdot \left( -\E[\mathrm{editing}|X,S,\phi,A=\mathrm{accept}] + \E[\mathrm{writing}|X,S, \phi, A=\mathrm{reject}] \right) \\
    &=  \E[\mathrm{verification}|X,S,\phi]  +\E[\mathrm{writing}|X,S, \phi, A=\mathrm{reject}]  - \E[\mathrm{writing}|X,\phi] + \E[\tau|X,\phi] 
\end{align*}

Assuming that $\E[\mathrm{writing}|X,S, \phi, A=\mathrm{reject}] - \E[\mathrm{editing}|X,S,\phi,A=\mathrm{accept}]$ is $
>0$, meaning when suggestions are accepted, the editing time for them is less  than the time to write the suggestions. 
We can then  separate $\bP^*(X,S,\phi)$ to the LHS with full equivalence (equality still holds without this assumption):
\begin{align*}
    & \bP^*(X,S,\phi) = \frac{\E[\mathrm{verification}|X,S,\phi]  + \E[\mathrm{writing}|X,S, \phi, A=\mathrm{reject}] -  \E[\mathrm{writing}|X,\phi] + \E[\tau|X,\phi]   }{   \E[\mathrm{writing}|X,S, \phi, A=\mathrm{reject}] - \E[\mathrm{editing}|X,S,\phi,A=\mathrm{accept}] } 
\end{align*}
Note that $\bP^*(X,S,\phi)$ is not necessarily a valid probability in $[0,1]$. If we assume that if the programmer rejected a suggestion, they did not benefit at all from it when writing the code afterwards meaning $\E[\mathrm{writing}|X,S, \phi, A=\mathrm{reject}] =  \E[\mathrm{writing}|X,\phi]  $, we arrive at: 

\begin{align*}
    & \bP^*(X,S,\phi) = \frac{\E[\mathrm{verification}|X,S,\phi]  + \E[\tau|X,\phi]  }{   \E[\mathrm{writing}|X,S, \phi, A=\mathrm{reject}] - \E[\mathrm{editing}|X,S,\phi,A=\mathrm{accept}] } 
\end{align*}
Assuming latency is zero meaning $\tau = 0$, then
it is clear that $\bP^*(X,S,\phi) =0$ if and only if  verification time is negligible: $\E[\mathrm{verification}|X,S,\phi] =0$, and by assumption it is $ \bP^*(X,S,\phi) \geq 0$ since the denominator is positive. 
The implication is that if $\bP(A=\textrm{accept}|X,S,\phi) \leq \bP^*(X,S,\phi)$, we should not show the suggestion which follows directly from the equation of $\delta \leq 0$, and similarly if $\bP(A=\textrm{accept}|X,S,\phi) \geq \bP^*(X,S,\phi)$ we show the suggestion.

To restate, this derivation made the following assumptions:
\begin{itemize}
    \item The programmer has only two actions of accept or reject.
    \item $\E[\mathrm{writing}|X,S, \phi, A=\mathrm{reject}] - \E[\mathrm{editing}|X,S,\phi,A=\mathrm{accept}]$ is $> 0$, meaning when suggestions are accepted, the editing time for them is less  than the time to write the suggestions.
    \item If we assume that if the programmer rejected a suggestion, they did not benefit at all from it when writing the code afterwards meaning $\E[\mathrm{writing}|X,S, \phi, A=\mathrm{reject}] =  \E[\mathrm{writing}|X,\phi]  $.
\end{itemize}

\end{proof}

\section{Model Evaluation and Analysis}\label{apx:experiments}

All experiments were run with Python 3.8 on a machine with a single A100 GPU.

\noindent Confidence intervals in the body are obtained by bootstrapping with 1000 bootstrap samples for accuracy and AUC.

In table \ref{tab:baselines} we evaluate the performance of different models for predicting acceptance of suggestions.
We use Scikit-learn \cite{scikit-learn} to train the Random Forrest and logistic regression models, XGBoost library to train our XGB models \cite{Chen:2016:XST:2939672.2939785} and PyTorch to train the fully connected neural network \cite{NEURIPS2019_9015}. The FCNN is a 3 layer network with 50 hidden units trained using Adam with a learning rate of $1e-2$. We used the validation set to select the number of hidden units as well as the picking the best model after training for 100 epochs. We did not employ any hyperparameter tuning for the XGBoost models and used the standard parameters.

\begin{table}[H]
\caption{Comparison of different classifiers on the test set based on AU-ROC, accuracy, and macro f1 for the full model to predict programmer suggestion acceptance. FCNN refers to a fully connected neural network.}
\centering
    \resizebox{0.48\textwidth}{!}{

\begin{tabular}{lccc}
    \toprule
    \textbf{Method} & \textbf{AU-ROC} & \textbf{Accuracy (\%)}  & \textbf{Macro F1 (\%)} \\
    \midrule
    \textbf{XGBoost} & 0.780 & 81.1 & 64 \\
    Logistic Regression & 0.726 & 79.5 & 58\\
    Random Forrest & 0.756 & 80.9 & 61 \\
    FCNN & 0.741  & 80.2 & 59\\
    Baseline & 0.5 & 78.9 & 44 \\
  \bottomrule
\end{tabular}}
    \label{tab:baselines}
\end{table}

\begin{figure}[H]
    \centering
    \includegraphics[scale=0.5]{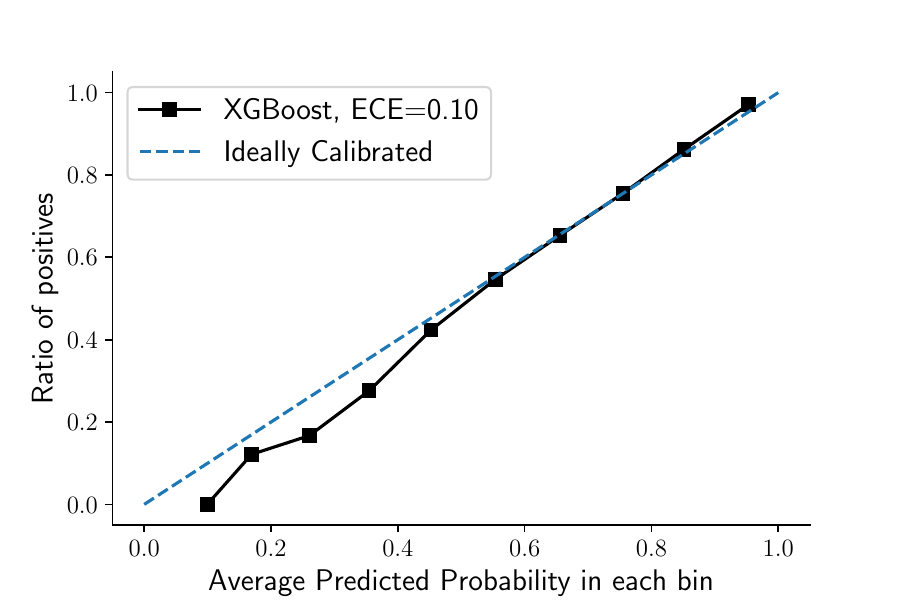}
        \caption{Calibration curve for the  XGBoost stage 2 model.}
    \label{fig:calibration_curve}
\end{figure}

In figure \ref{fig:calibration_curve} we showcase that the stage 2 model $m_2$ is well calibrated.

\begin{figure}[H]
    \centering
    \includegraphics[scale=0.6]{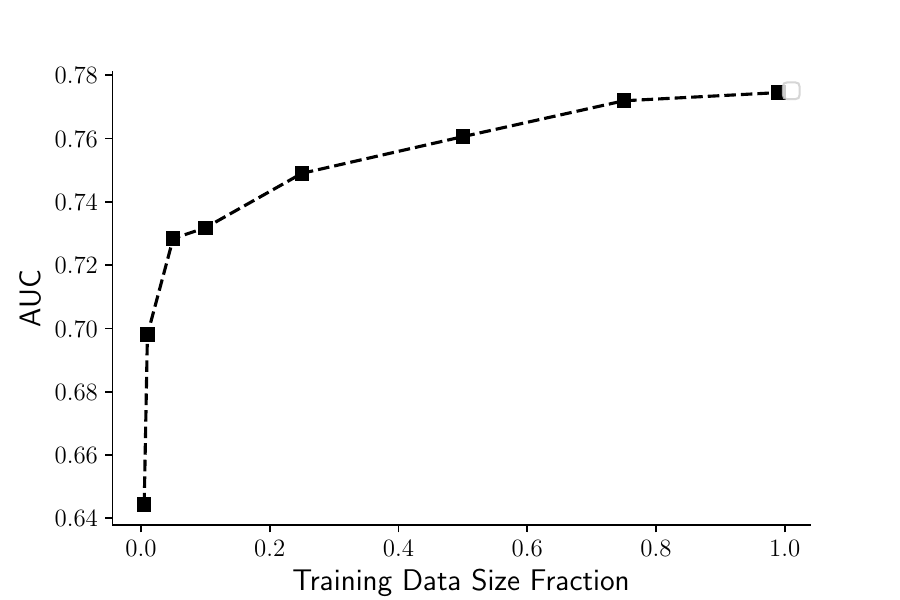}
        \caption{Sample complexity analysis of the XGBoost stage 2 model when trained on a fraction of the training data and plotting the AU-ROC on the full test set.}
    \label{fig:learning_curveapx}
\end{figure}

\textbf{Sample Complexity.} To understand how many samples we might need to train individualized CDHF models, we perform a sample complexity analysis on the acceptance prediction stage 2 model where we train with a random fraction of the training dataset in Figure \ref{fig:learning_curveapx}. With 1\% of the training dataset which equates to 1688 training samples, performance reaches 0.69 AU-ROC. With 25\% of training data, the model can achieve 0.75 AU-ROC.

\begin{figure}[H]
    \centering
    \includegraphics[ width=0.5\textwidth]{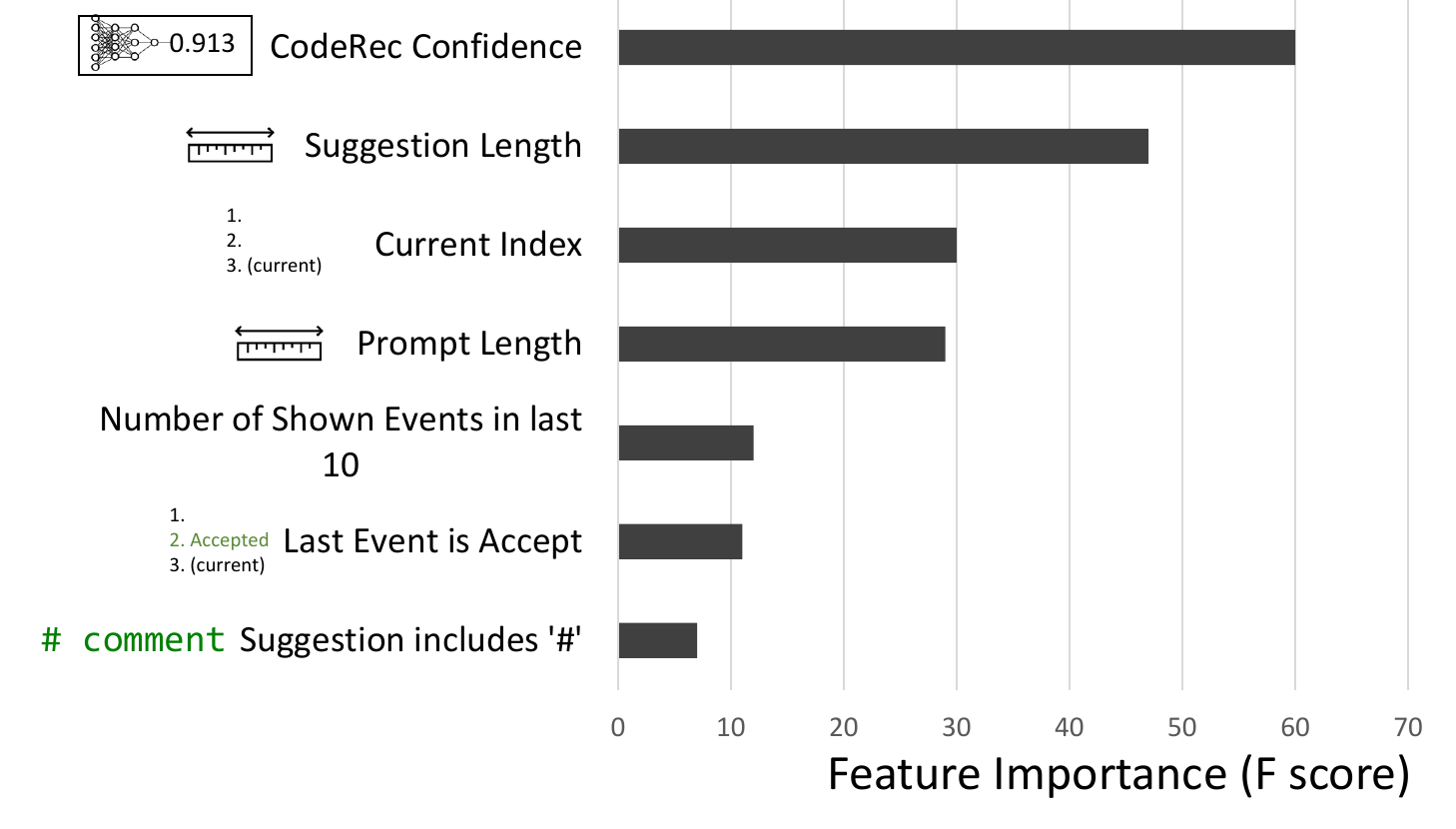}
    \caption{Feature importance for the seven highest-rated unique features of the model for predicting the likelihood of accepting a suggestion. The feature importance is in terms of the F score which counts how often a feature was split on in the tree ensemble. }
    \label{fig:feature_importance}
\end{figure}

\textbf{Factors Influencing Programming Actions.} An analysis of the XGBoost stage 2 model, reveals factors that correlate with programmers' decisions. We examine feature importance weights (Figure \ref{fig:feature_importance}). Specifically, we report feature F-score counts, which capture how often each feature was split in any of the trees in the XGBoost model. The two most important features are the confidence of the suggestion from \coderec's  core suggestion model, and the length of the suggestion. Together, these two features yield a model that achieves an AUROC of 0.71. Other important features include the context of the current event in the coding session such as if the last suggestion was accepted or not. We also see that textual elements of the suggestion are important. For example, the feature indicating if the current suggestion includes the character '\#' (which is used to indicate a comment in Python) was split a total of 8 times. These features should not be interpreted in a causal fashion but rather as being correlated with the behavior of programmers.

\textbf{Analysis of Suggestions.} The model learned on the large sample of telemetry can also be applied to the telemetry collected in the 21-participant user study of \cite{mozannar2022reading}. When we evaluate the model on the user study's 1029 accepted and rejected events we obtain an AU-ROC of 0.73. 
Inspecting these results further, we find that there are at least two defined clusters for suggestions predicted most likely be rejected: (1) single character non-alphabetic suggestions such as {\tt (}, {\tt )}, {\tt [},{\tt :}, {\tt ;}, and (2) mid-word completions such as `{\tt agonal()}' (completion of `{\tt diagonal()}'), `{\tt lass LogisticRegression:}' (completion of `{\tt class Logistic Regression:}'. We hypothesize that for  cluster (1) the suggestions were too small to be noticed. For cluster (2), we hypothesize that the programmer was already in the act of typing, so the suggestions may have been a distraction (i.e., the interruption cost more than was saved by eliminating the physical act of typing a few already-determined keystrokes).

\section{Which Suggestion to Show: Plots}

Following the last section of the paper, we show two plots for each $k \in \{0,1,2,3\}$: 1) histogram showing in which length percentile in terms the ground truth solution dos the suggestion with highest acceptance probability (according to the model) lies and 2) normalized probability of acceptance by the length of the suggestion (for each example, we normalize the raw probability of acceptance by the maximum acceptance probability across all length for the given example - we observe the same trend without normalizing).

\begin{figure}[H]
  \centering
  \subfigure[Histogram of position of max suggestion]{
    \includegraphics[width=0.45\textwidth]{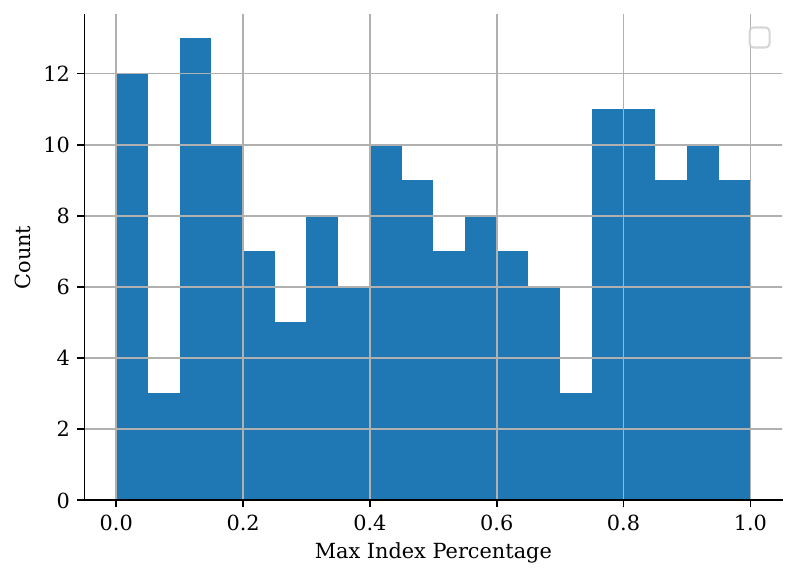}
    \label{fig:plot1}
  }
  \hfill
  \subfigure[Probability of acceptance by length]{
    \includegraphics[width=0.45\textwidth]{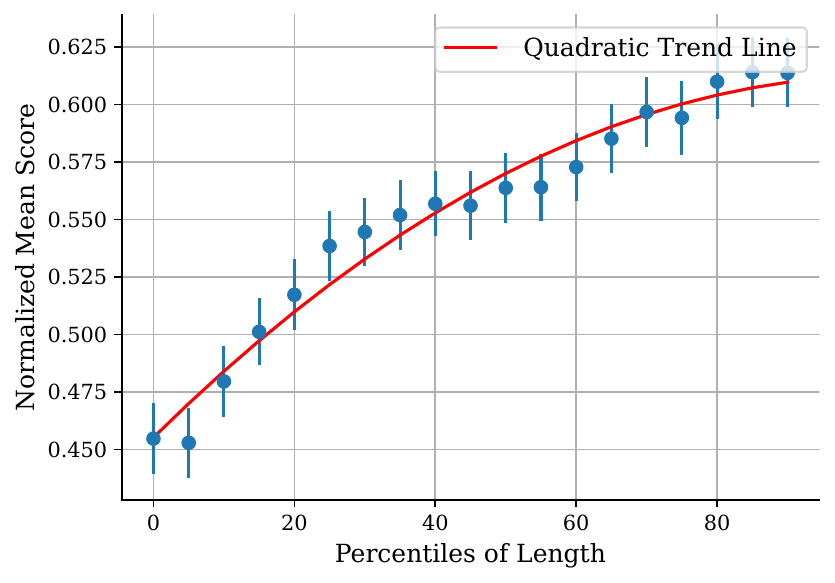}
    \label{fig:plot2}
  }
  \caption{Plots for the experiment on ranking suggestions by the probability of acceptance. Histogram (a) shows in which length percentile bin the maximizing suggestion lies and Graph (b) shows the acceptance score by increasing the length of the suggestion. These plots are for $k=0$ (docstring only) }
  \label{fig:which_k0}
\end{figure}

\begin{figure}[H]
  \centering
  \subfigure[Histogram of position of max suggestion]{
    \includegraphics[width=0.45\textwidth]{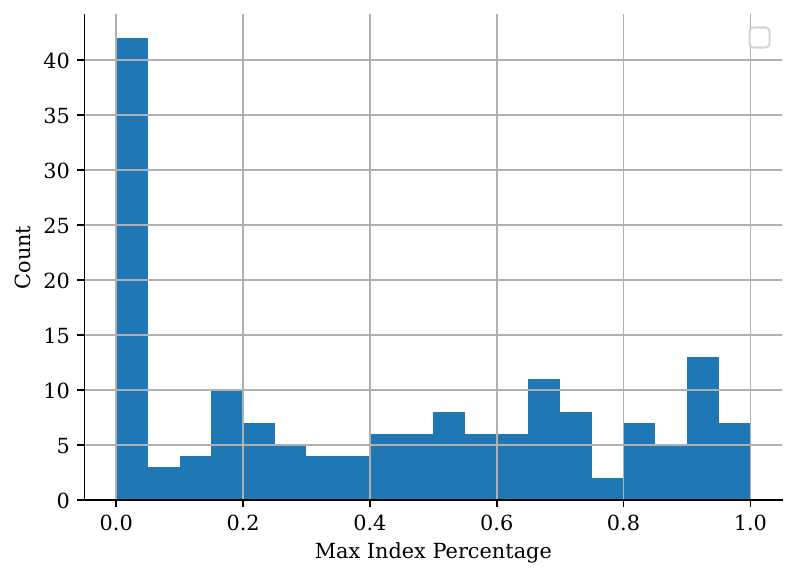}
    \label{fig:plot1}
  }
  \hfill
  \subfigure[Probability of acceptance by length]{
    \includegraphics[width=0.45\textwidth]{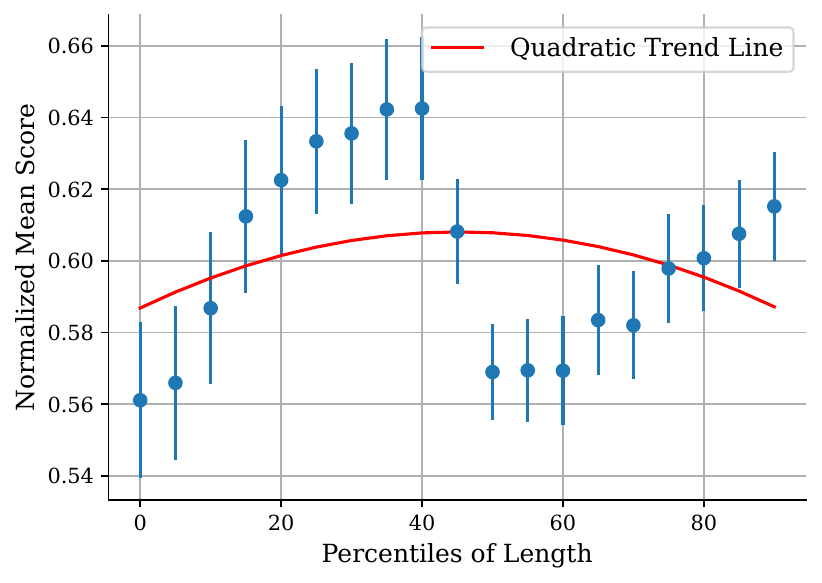}
    \label{fig:plot2}
  }
  \caption{Plots for the experiment on ranking suggestions by the probability of acceptance. Histogram (a) shows in which length percentile bin the maximizing suggestion lies and Graph (b) shows the acceptance score by increasing the length of the suggestion. These plots are for $k=1$ (docstring + first line of solution) }
  \label{fig:which_k1}
\end{figure}

\begin{figure}[H]
  \centering
  \subfigure[Histogram of position of max suggestion]{
    \includegraphics[width=0.45\textwidth]{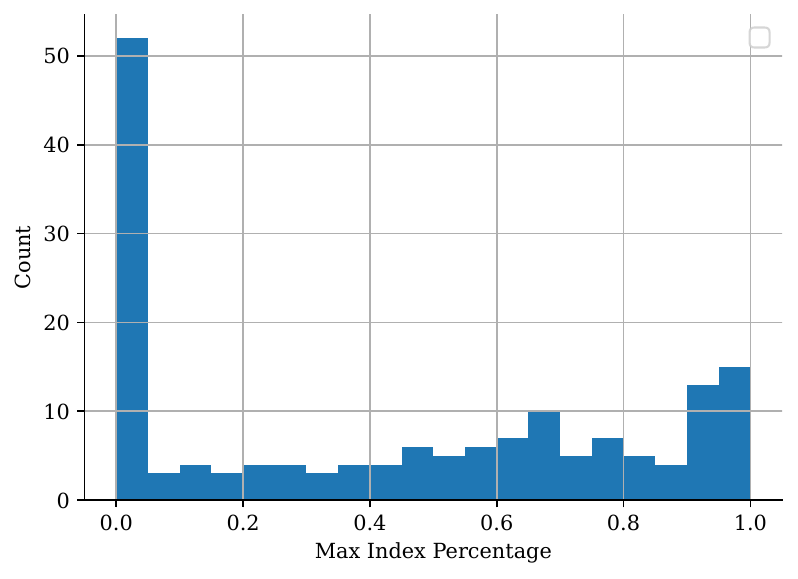}
    \label{fig:plot1}
  }
  \hfill
  \subfigure[Probability of acceptance by length]{
    \includegraphics[width=0.45\textwidth]{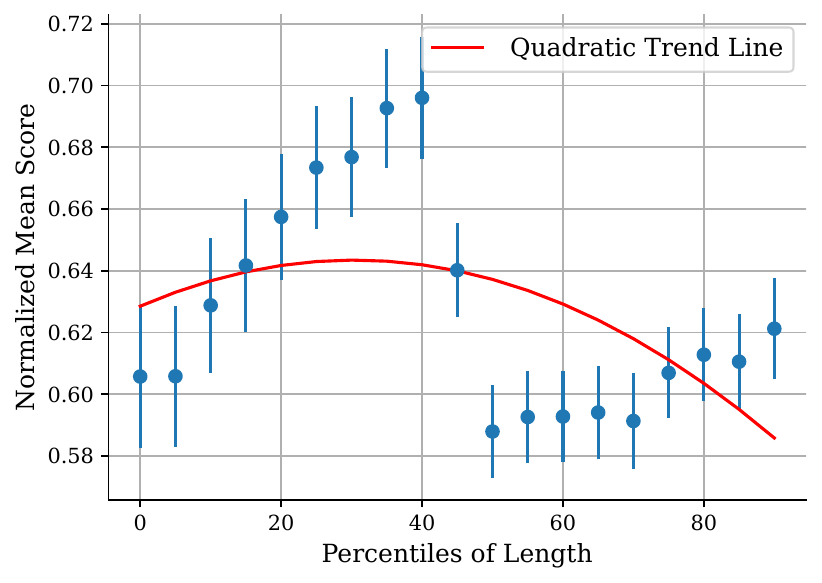}
    \label{fig:plot2}
  }
  \caption{Plots for the experiment on ranking suggestions by the probability of acceptance. Histogram (a) shows in which length percentile bin the maximizing suggestion lies and Graph (b) shows the acceptance score by increasing the length of the suggestion. These plots are for $k=2$ (docstring + first two lines of solution) }
  \label{fig:which_k2}
\end{figure}

\begin{figure}[H]
  \centering
  \subfigure[Histogram of position of max suggestion]{
    \includegraphics[width=0.45\textwidth]{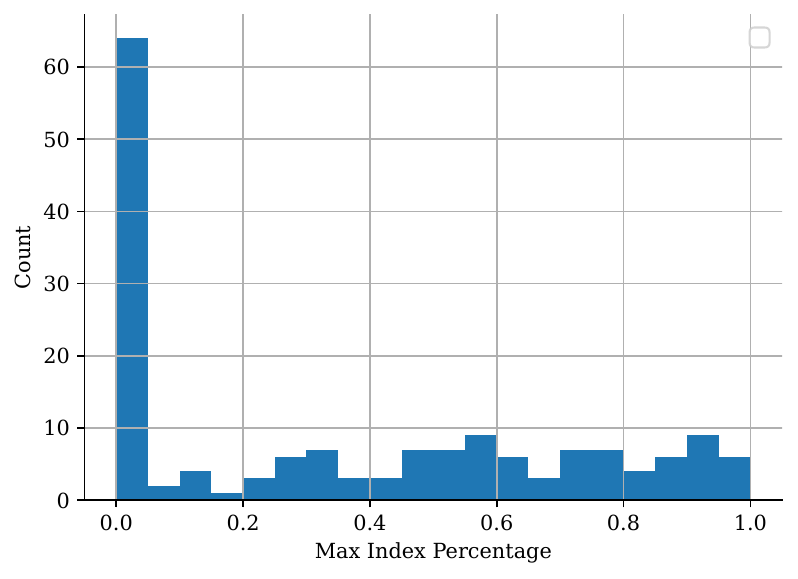}
    \label{fig:plot1}
  }
  \hfill
  \subfigure[Probability of acceptance by length]{
    \includegraphics[width=0.45\textwidth]{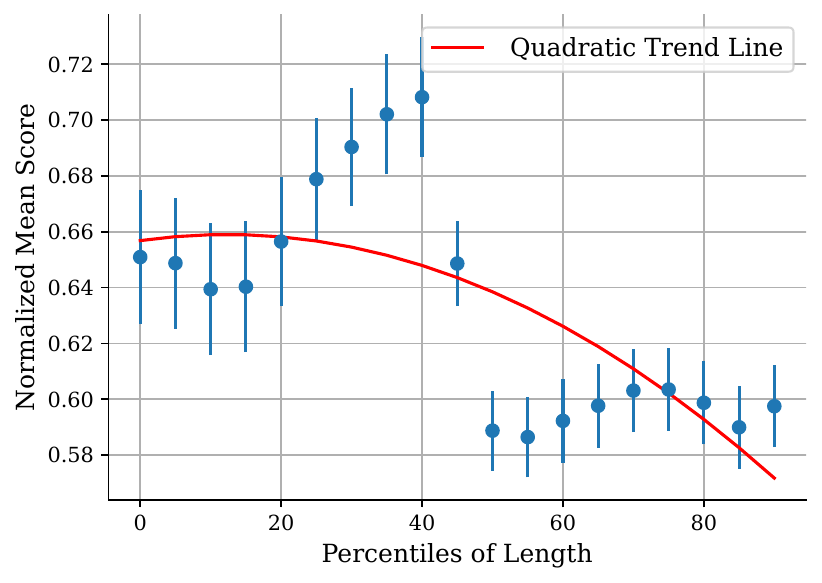}
    \label{fig:plot2}
  }
  \caption{Plots for the experiment on ranking suggestions by the probability of acceptance. Histogram (a) shows in which length percentile bin the maximizing suggestion lies and Graph (b) shows the acceptance score by increasing the length of the suggestion. These plots are for $k=3$ (docstring + first three lines of solution) }
  \label{fig:which_k3}
\end{figure}

\end{document}